\documentclass[preprint,12pt]{elsarticle}
\usepackage{amssymb}
\usepackage{times}
\usepackage{xcolor}
\usepackage{soul}
\usepackage[utf8]{inputenc}
\usepackage[small]{caption}
\usepackage{amsmath}
\usepackage{paralist}
\usepackage{graphicx}
\usepackage{multirow}
\usepackage{url}
\journal{**}

\begin{document}

\begin{frontmatter}

\title{Operation-aware Neural Networks for User Response Prediction}
\author[nju]{Yi Yang}

\author[nju]{Baile Xu}

\author[nju]{Furao Shen\corref{mycorrespondingauthor}}
\cortext[mycorrespondingauthor]{Corresponding authors}
\ead{frshen@nju.edu.cn}

\author[nju2]{Jian Zhao\corref{mycorrespondingauthor}}
\ead{jianzhao@nju.edu.cn}

\address[nju]{State Key Laboratory for Novel Software Technology,\\
Department of Computer Science and Technology,\\
Collaborative Innovation Center of Novel Software Technology and Industrialization,\\
Nanjing University, China}
\address[nju2]{State Key Laboratory for Novel Software Technology,\\
    School of Electronic Science and Engineering, Nanjing University, China\\}

\begin{abstract}
User response prediction makes a crucial contribution to the rapid development of online advertising system and recommendation system.
The importance of learning feature interactions has been emphasized by many works.
Many deep models are proposed to automatically learn high-order feature interactions.
Since most features in advertising system and recommendation system are high-dimensional sparse features, deep models usually learn a low-dimensional distributed representation for each feature in the bottom layer.
Besides traditional fully-connected architectures, some new operations, such as convolutional operations and product operations, are proposed to learn feature interactions better.
In these models, the representation is shared among different operations.
However, the best representation for different operations may be different.
In this paper, we propose a new neural model named \emph{Operation-aware Neural Networks} (ONN) which learns different representations for different operations.
Our experimental results on two large-scale real-world ad click/conversion datasets demonstrate that ONN consistently outperforms the state-of-the-art models in both offline-training environment and online-training environment.
\end{abstract}


\end{frontmatter}


\section{Introduction}
In recent years, online advertising develops rapidly among social medias such as Facebook and Wechat.
As a critical role of online advertising, user response prediction makes a crucial contribution, where the task is to estimate the probability of a user will click on an ad (click-through rate, CTR) or take a desired action after clicking the ad (conversion rate, CVR).

In CTR/CVR prediction tasks, the importance of learning feature interactions has been emphasized by many works in related literature \cite{rendle2010factorization,juan2016field}.
For instance, users of different ages have different preferences for different types of ads, which suggests that the interaction between \emph{User Age} and \emph{Ad Type} is a strong signal for CTR/CVR prediction.
Traditional linear models with manual feature engineering have shown decent results, but manual feature engineering is very labor-intensive and time-consuming.
At the same time, it is very difficult for human experts to discover high-order interactions between features.

Deep neural networks (DNN) show very promising results at automatically learning feature representations and dependencies.
Driven by the success of DNN, several neural architectures are proposed for CTR/CVR prediction in recent years.
Most of these architectures can be concluded into 3 steps:

\begin{inparaenum}[\itshape i\upshape)]
\item An embedding layer is used to map high-dimensional sparse features into low-dimensional distributed representations.
The output of the step is $e = [V^{0}x_{0}, V^{1}x_{1}, \cdots , V^{m}x_{m}]$, where $V^{i}$ is the embedding matrix of the $i$th feature and $x_{i}$ is the one-hot representation of the $i$th feature.

\item \label(step2) Several operations are applied on the embedding vectors to get the medial features.
The output of this step is $f = [o_{1}(e), o_{2}(e), \cdots , o_{l}(e)]$ where $o_{i}$ is the $i$th operation.
In most architectures, the operation is just a copy of the embedding vectors.

\item A multi-layer perceptron (MLP) is applied on $f$ to learn nonlinear relations among features.
The output of this step is $\widehat{y} = \sigma(\Phi(f))$ where $\sigma$ is the \emph{sigmoid} function and $\Phi$ is the multi-layer non-linear transformation.
Some architectures also have some other components.
In these models, the output is $\widehat{y} = \sigma(\Phi(f) + \Delta)$ where $\Delta$ denotes the specific components.
\end{inparaenum}

Some recent works focus on introducing new operations to learn the feature interactions better, such as convolutional operations \cite{liu2015convolutional} and ``inner-product/outer-product'' operations \cite{qu2016product}.
Different operations play different roles in deep models.
For instance, each ``copy" operation reserves the original embedded representation for one feature.
Each ``inner-product" operation or ``outer-product" operation learns a local dependency between two features.
The operations on embedding vectors can be regarded as an incipient feature engineering before applying MLP.

On the other hand, the feature representation is very important to the model because a better representation makes learning feature interactions easier.
However, few works focus on improving the feature representation learned by the embedding layer.
Existing models usually share the same feature representation among different operations.
However, the best feature representations for different operations are not always the same.
It has been experimentally proven in previous works \cite{rendle2010pairwise,juan2016field} that structures explicitly using different embeddings among different operations perform better than structures sharing one embedding among all operations in many tasks.

Inspired by this idea, we propose a new embedding method named \emph{operation-aware} embedding in this paper, which learns different representations for each feature when performing different operations.
Models with operation-aware embedding layer are named as\emph{ Operation-aware Neural Networks} (ONN).
The operation-aware embedding makes ONN more flexible than exsiting models.
Compared with state-of-the-art models, ONN shows superior performances in many CTR/CVR tasks.
Furthermore, experimental results show that ONN is especially suitable for online-training environments.
We have used ONN to win the first prize of Tencent Social Advertising College Algorithm Competition among about 1000 teams\footnote{http://algo.tpai.qq.com/home/home/index.html}.


The rest of the paper is organized as follows. Related works are introduced in section II. The model details are described in Section III. We discuss feature embedding and the relationships of ONN with related models in Section IV. Section V exhibits experiments analysis. Finally, we conclude the paper in Section VI.

\section{Related Works}
The CTR/CVR tasks, which can be formalized as a classic binary classification problem, have been intensely studied in the literature.
Logistic Regression with FTRL optimizer \cite{mcmahan2013ad} has been widely used in real world applications.
However, the linear model needs a lot of artificial feature engineering to generate the feature interactions.
In order to solve this problem, Factorization Machines (FM) \cite{rendle2010factorization} are proposed to learn the feature interactions automatically.

With the rapid development of deep learning, many deep methods are also proposed to solve the CTR/CVR tasks.
Factorization-machine supported Neural Network (FNN) is the first deep model for CTR/CVR tasks, which uses a MLP to learn high-order feature dependencies on the hidden vectors of FM.
Without FM initialization, FNN is equivalent to a standard MLP above an embedding layer.
For the purpose of learning better feature interactions, Convolutional Click Prediction Model (CCPM) \cite{liu2015convolutional}, DeepCross \cite{Wang2017DeepCross} and Product-based Neural Network (PNN) \cite{qu2016product} introduce some new operations to be performed on the embedded features before applying MLP.
Some researchers think that the shallow component can be a supplement to the deep models.
For instance, the Wide $\&$ Deep model of Google \cite{cheng2016wide} uses a linear component as the supplement to the deep component.
Similarly, deepFM \cite{guo2017deepfm} is proposed by training a FM component and a deep component at the same time.
Besides, there are also some researches which focus on improving the model performance by mining the important features.
For instance, Deep Interest Network (DIN) \cite{zhou2018deep} proposes using attention mechanism to adaptively learn the representation of user interests from historical behaviors.
The attention mechanism is also used in \cite{xiao2017attentional}, but it is used to learn the weights of feature interactions.

To the extent of our knowledge, few works focus on improving the embedding layer.
In following parts of the paper, we demonstrate the operation-aware embedding can improve the model performance by learning better feature representations.
\section{Operation-aware Neural Networks}
In an advertising system, each datum consists of many features including user information (Age, Education, etc.), ad information (Publish Company, Ad Type, etc.), context information (Connection Type, Siteset ID , etc.).
We only consider the case of categorical features here because most features in ad systems are either categorical or can be made categorical through discretization.
Each feature is represented as a vector of one-hot encoding after data preprocessing.
Suppose each datum have $m$ features, then each training sample can be represented as $(x, y)$ where $x=[x_{1}, x_{2}, ... , x_{m}]$ with $x_{i}$ standing for the one-hot vector of the $i$th feature and $y \in \{0, 1\}$ records whether the user performed a positive action.
The objective of a user response prediction system is learning a function $f:x \rightarrow \widehat{y}$ that maps the input features $x$ to the estimated probability of positive response $\widehat{y} \in [0, 1]$.
\begin{figure}
\centering
\includegraphics[width=8.5cm]{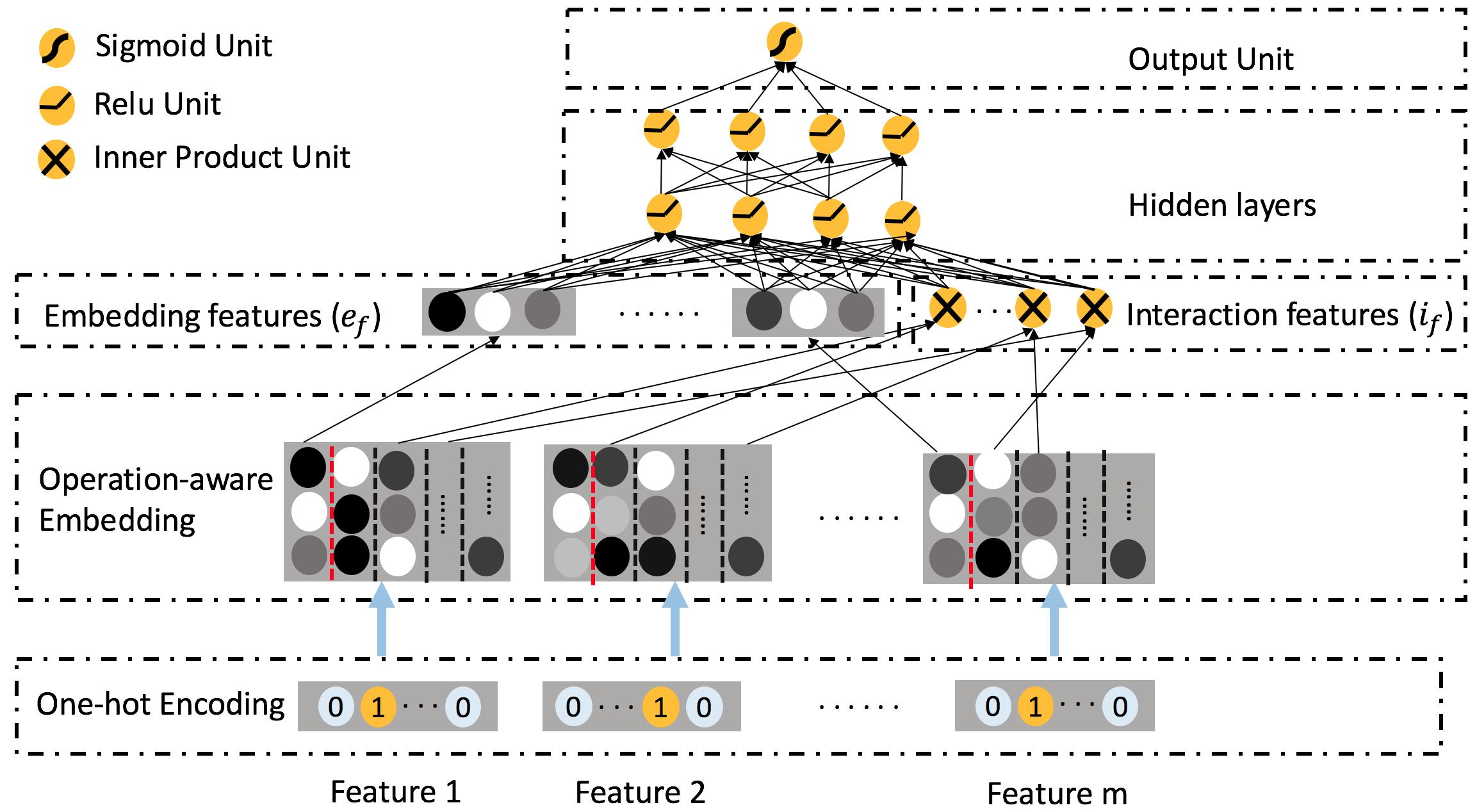}
\caption{
The architecture of the ONN model.
From the bottom-up perspective, the model first uses an operation-aware embedding layer to map high-dimensional sparse features into low-dimension representations.
After that, ``copy" operations and ``inner-product" operations are performed on corresponding features respectively.
Finally, an MLP learns high-order feature interactions and produce the output.}
\label{ONN_pic}
\end{figure}

\subsection{Model Architecture Design}
The architecture of the ONN model is illustrated in Figure \ref{ONN_pic}.
From the bottom-up perspective, the model can be divided into 3 components:
\subsubsection{Operation-aware Embedding}
The one-hot encoding is inefficient and can not represent correlations among features.
In addition, many operations are meaningless with one-hot encoding, such as the ``inner-product" operation.
In order to solve this, we use an embedding layer to map the high-dimensional one-hot vectors into low-dimensional vectors.
Compared with one-hot encoding, embedded representation is more efficient, informative, and learnable.

Existing models usually learn one representation for each feature and use the same representation among all operations.
However, the best representation may be different among different operations.
Why can one feature have just one representation?
If we use one representation for all operations, the representation needs to compromise among all operations.
On the bright side, this compromise might have the effect of regularization sometimes, especially when the training data are not sufficient.
But in CTR/CVR tasks where training data are easy to obtain, the compromise tends to limit the expression ability of the model.

To learn different representations for different operations, we propose the \emph{operation-aware embedding} method.
For each feature, an embedding vector is learned for each operation performed on it.
Note that \emph{operations of the same type performed on different features are also regarded as different operations}.
For instance, for feature $i$, the ``inner-product" with feature $j$ and the ``inner-product" with feature $p$ are regarded as different operations.
The similar idea is also used in Pairwise Interaction Tensor Factorization (PITF) \cite{rendle2010pairwise} and Field-aware Factorization Machine (FFM) \cite{juan2016field}.

\begin{figure}
\centering
\includegraphics[width=8.5cm]{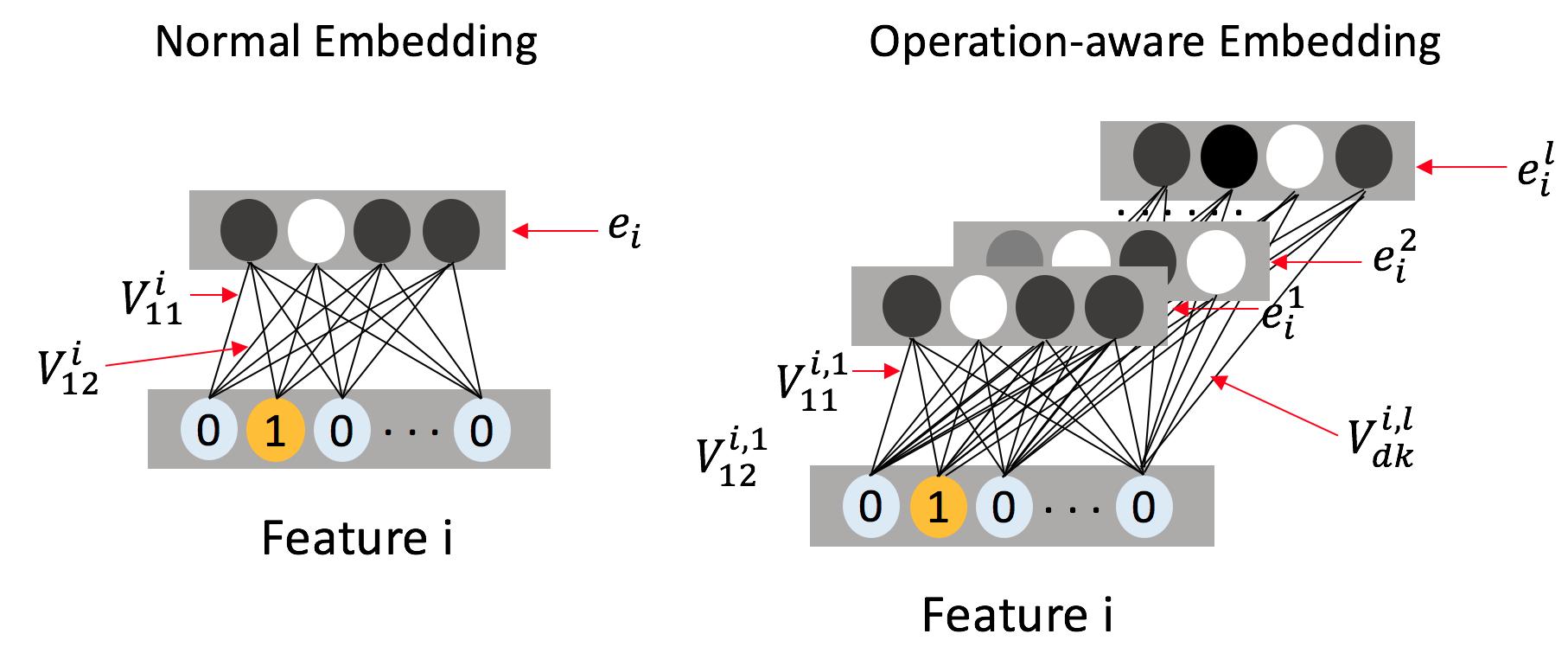}
\caption{The structures of the normal embedding layer and the operation-aware embedding layer.}
\label{OE_pic}
\end{figure}

Figure \ref{OE_pic} illustrates the difference between the normal embedding layer and the operation-aware embedding layer.
In the normal embedding layer, the embedded representation for all operations of the $i$th feature is $e^{i} = V^{i}x_{i}$ where $V^{i}$ is embedding matrix of the $i$th feature.
However, in the operation-aware embedding layer, each feature has several embedded representations.
Suppose there are $l$ operations performed on the $i$th feature, we use $[V^{i,1}, V^{i,2}, \cdots ,  V^{i,l}]$ to denote the embedding matrices of the $i$th feature.
Then the embedded representations of the $i$th feature are $[e_{i}^{1}, e_{i}^{2}, \cdots , e_{i}^{l}]=[V^{i,1}x_{i}, V^{i,2}x_{i}, \cdots , V^{i,l}x_{i}]$.
When performing the $k$th operation on the $i$th feature, $e_{i}^{k}$ is used as the representation to perform the computation.

In order to use \emph{operation-aware} representation for each operation, we need to construct a mapping $\Gamma$ which maps an operation and a feature index to the operation index of the feature.
Specifically speaking, $\Gamma(o, i)=k$ means that the operation $o$ is the $k$th operation to be done on the $i$th feature.

\subsubsection{Incipient Feature Extraction Layer}
Besides the embedded representations for the raw features, we also hope that the bottom layer can provide representations for the feature interactions because the feature interactions are very important in CTR/CVR tasks. Intuitively, 
we can use the one-hot encoding to encode the interaction of the two features and then embed it into a low-dimensional vector.
However, the direct interaction of two features usually leads to a very sparse feature whose representation can not obtain adequate training.
Factorization Machines (FM) solves the problem by factorizing the effect of feature interactions into a product of two latent vectors.
In the PNN model, \cite{qu2016product} solves the problem in the same way.
They use the ``inner-product" operation and the ``outer-product" operation of the embedding vectors to represent the feature interactions.
These works suggest that ``product" operations are effective in learning feature interactions.

We use the same idea in the ONN model.
Because the experimental results in \cite{qu2016product} demonstrate that the ``inner-product" operation performs better than the ``outer-product" operation most of time, we use the ``inner-product" operation as the default operation to learn the feature interactions.
We only consider 2-order interactions because higher-order interactions are too complicated.
We use $o(c, i)$ to denote the ``copy" operation of the $i$th feature, and $o(p, i, j)$ to denote the ``inner-product" operation between the $i$th feature and the $j$th feature.
The output of this layer can be split into the embedded features $e_{f}$ and the interaction features $i_{f}$.
Suppose we have $m$ features, $e_{f}$ can be constructed as follow:
\begin{equation}
e_{f} = [e_{1}^{\Gamma(o(c,1), 1)}, e_{2}^{\Gamma(o(c,2), 2)}, \cdots , e_{m}^{\Gamma(o(c,m),m)}].
\end{equation}
recall that $\Gamma$ is the mapping from operation to index. $i_{f}$ can be constructed as follow:
\begin{equation}
i_{f} = [p_{1,2}, p_{1,3}, \cdots, p_{m-1,m}]
\end{equation}
where $p_{i,j}$ is the value of the ``inner-product" operation between the $i$th feature and the $j$th feature:
\begin{equation}
p_{i,j}=<e_{i}^{\Gamma(o(p,i,j),i)}, e_{j}^{\Gamma(o(p,i,j),j)}>.
\end{equation}
Concatenating $e_{f}$ and $i_{f}$ constructs the output of this layer $f=[e_{f}, i_{f}]$.
\subsubsection{Multiple Nonlinear Layers}
Lastly, a multi-layer perceptron (MLP) is applied on the medial features $f$ to mining the sophisticated patterns among data and generate the model output.
For each layer of the MLP, we add a batch normalization (BN)\cite{ioffe2015batch} layer to accelerate the training.
We have conducted a large number of experiments and find that BN can not only greatly speed up the training, but also improve the prediction accuracy.
For the definition of BN, please refer to \cite{ioffe2015batch}.

We explain how to generate the output with an architecture consist of 2 hidden layers as an example.
Firstly we should do batch normalization on $f$.
Denote the result as $\widehat{f}=BN(f)$, then the output of the first hidden layer is:
\begin{equation}
l1 = BN(relu(W_{1}\widehat{f}+b_{1}))
\end{equation}
where $W_{1}$ and $b_{1}$ are the model weights and bias of the first hidden layer, $relu$ is the rectified linear unit \cite{nair2010rectified}, defined as $relu(x) = max(0,x)$.
Hence the output of the second hidden layer is:
\begin{equation}
l2 = BN(relu(W_{2}l_{1}+b_{2})).
\end{equation}
Then the output of the model is:
\begin{equation}
\widehat{y} = \sigma(W_{3}l_{2}+b_{3})
\end{equation}
where $\sigma$ is the sigmoid function, defined as $\sigma(x)=\frac{1}{1+exp(-x)}$.
After that, we minimize the log loss to train the model. The loss function is defined as:
\begin{equation}
L(y, \widehat{y}) = -ylog(\widehat{y}) - (1-y)log(1-\widehat{y}).
\end{equation}

\section{Discussions}
\subsection{Feature Embedding}
In natural language processing (NLP), a word may have various meanings in different contexts.
Multi-sense embeddings \cite{iacobacci2015sensembed,li2015multi} are proposed to solve the problem of one word having multiple meanings.
Similarly, each feature is faced with different contexts when performing different operations in CTR/CVR tasks.
The operation-aware embedding learns representations for each feature in different contexts.
The context adaptive representations are more informative then single representation, thus provide an improvement to the model performance.

Although we just use the ``copy" operation and the ``inner-product" operation in this paper, the ONN model can be generalized to more operations, such as the ``outer-product" operation mentioned in \cite{qu2016product}.
In fact, we have tried to use a sub-network as an operation type in our experiments and have also achieved pretty good results, but the time complexity is relatively higher.

The embedding layer in the ONN model is more flexible than that in the other models \cite{qu2016product,guo2017deepfm,zhang2016deep}, which share the same embedding among all operations and hence reduce their flexibility.
For example, the embedding dimensionality of User ID should be larger than that of User Education because the possible values of User ID are far more than that of User Education.
But if we share the same embedding between ``copy" operations and ``product" operations, the embedding dimension of User ID and User Education should be the same because ``product" operations need the 2 vectors to have the same length.
In contrast, we can use different embedding dimensionalities for different features in the ONN model.
For each feature, when performing product operations with different features, we can also use different embedding dimensionalities.
By artificially designing different embedding dimensionalities for each feature and each operation, we can minimize the computation complexity during training and testing.

\subsection{Relationships with Related Models}
\begin{figure}[h]
\centering
\includegraphics[width=8.5cm]{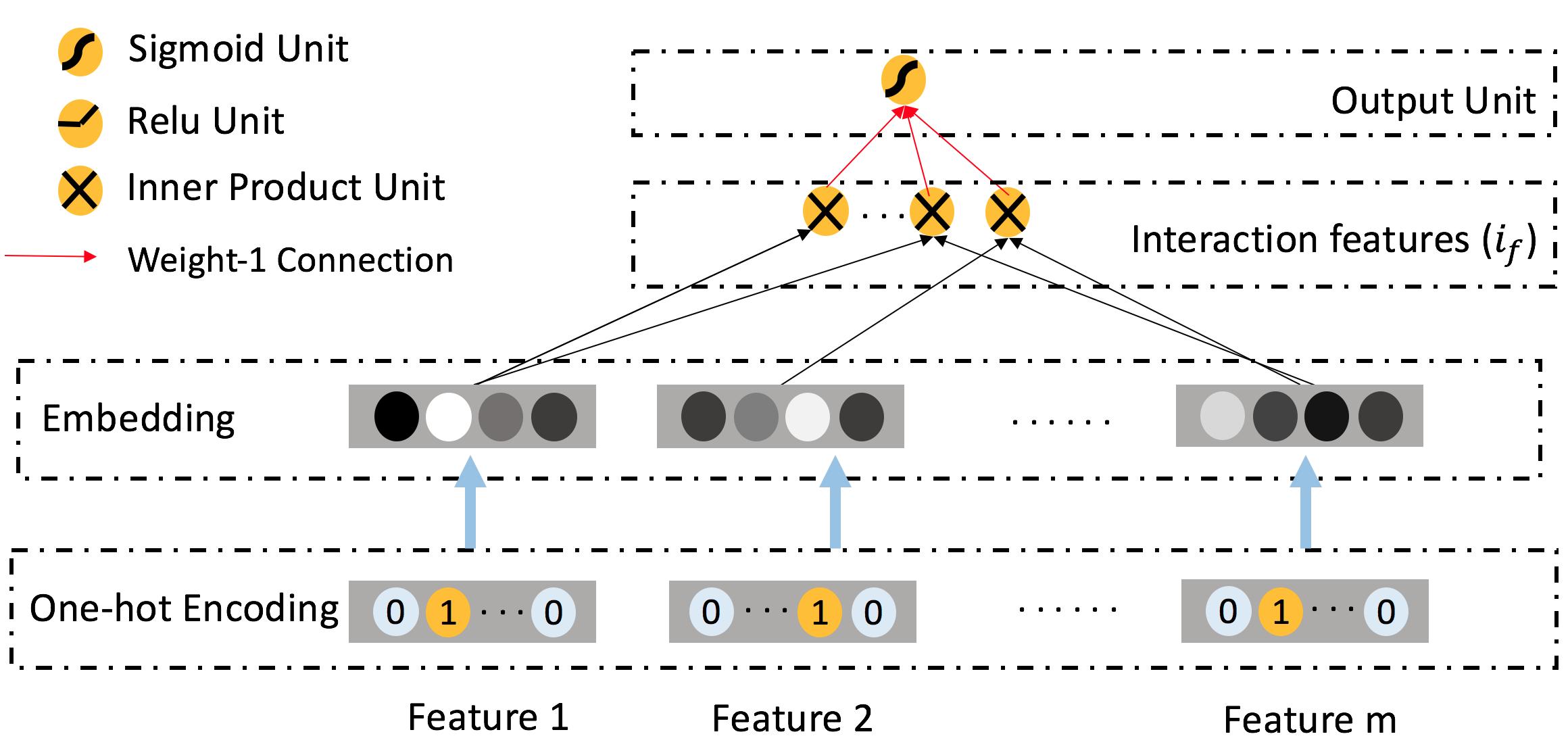}
\caption{The architecture of the FM model. The embedding vectors correspond to the latent vectors in FM. The sum of the inner product units is fed into the output unit.}
\label{FM_pic}
\end{figure}

\begin{figure}[h]
\centering
\includegraphics[width=8.5cm]{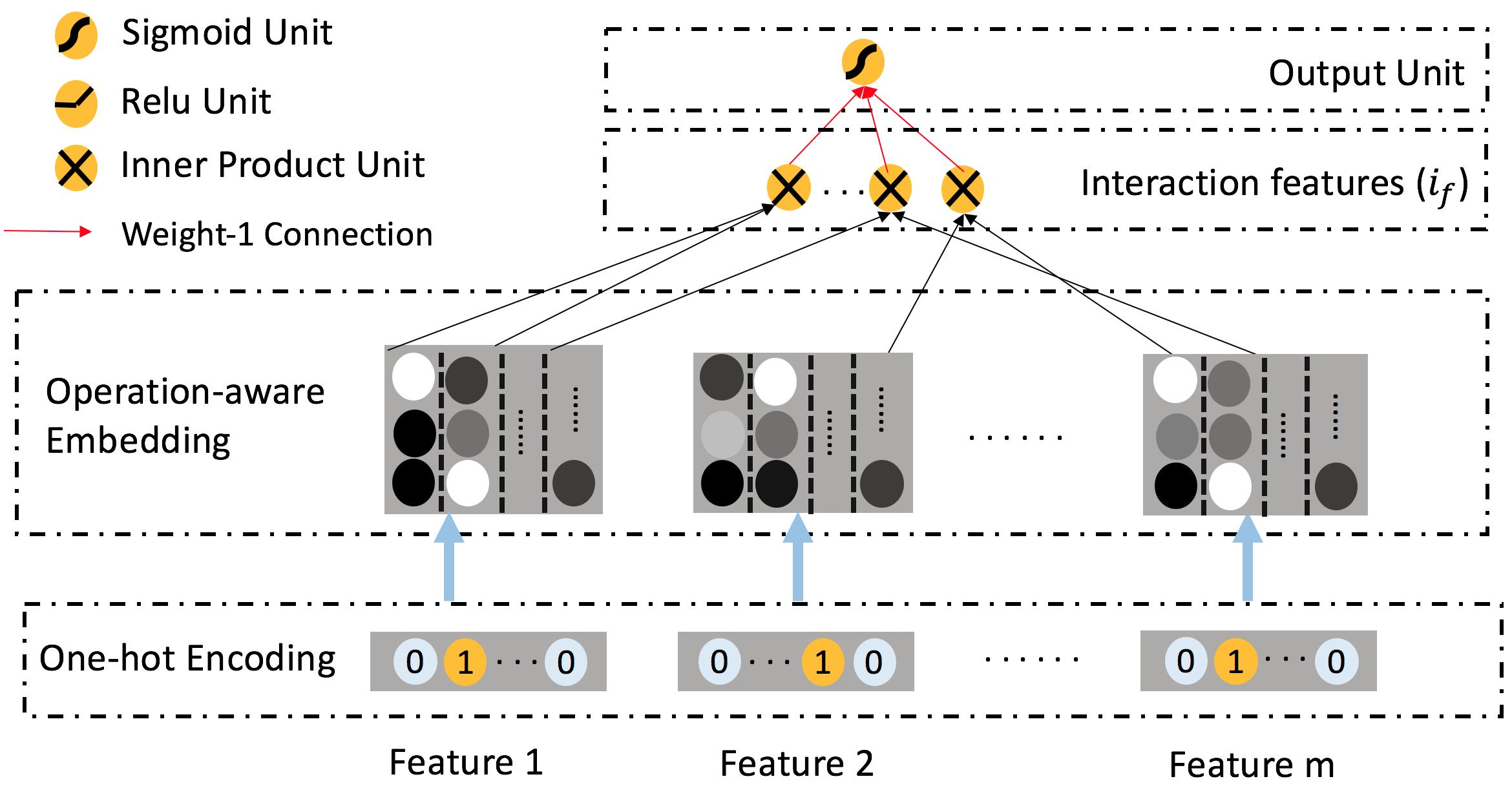}
\caption{The architecture of the FFM model. The structure is the same as FM, except that the embedding layer is operation-aware in FFM.}
\label{FFM_pic}
\end{figure}

In this subsection, we discuss the relationships of the ONN model and several related models.
Firstly, we give the neural architectures of FM and FFM.
If we only consider category features, the FM of degree 2 is defined as:
\begin{equation}
\phi_{fm}(w, x) = \Sigma_{i=1}^{m}\Sigma_{j=i+1}^{m}<v_{i},v_{j}>
\end{equation}
where $v_{i}$ is the latent vector of the $i$th feature.
Figure \ref{FM_pic} illustrates the structure of FM.
The bottom layer is a normal embedding layer, where each embedding vector corresponds to the latent vector of one feature in FM.
The embedding vectors are connected with the inner product units.
Each inner product unit corresponds to an inner product of 2 latent vectors in FM.
Finally, the product units are summed up to generate the output.
FFM is an optimized version of FM.
But when performing the product operation with different features, FFM uses field-aware vector for each feature.
The FFM model is defined as:
\begin{equation}
\phi_{ffm}(w, x) = \Sigma_{i=0}^{m}\Sigma_{j=i+1}^{m}<v_{i,j}, v_{j,i}>.
\end{equation}
Actually, ``field-aware" is a special case of ``operation-aware" when there is only the ``inner-product" operation.
Figure \ref{FFM_pic} illustrates the structure of FFM.
It can be seen that FM uses the normal embedding layer, but FFM uses the operation-aware embedding.
Thus, we say that FFM makes FM operation-aware.

\begin{figure}
\centering
\includegraphics[width=8.5cm]{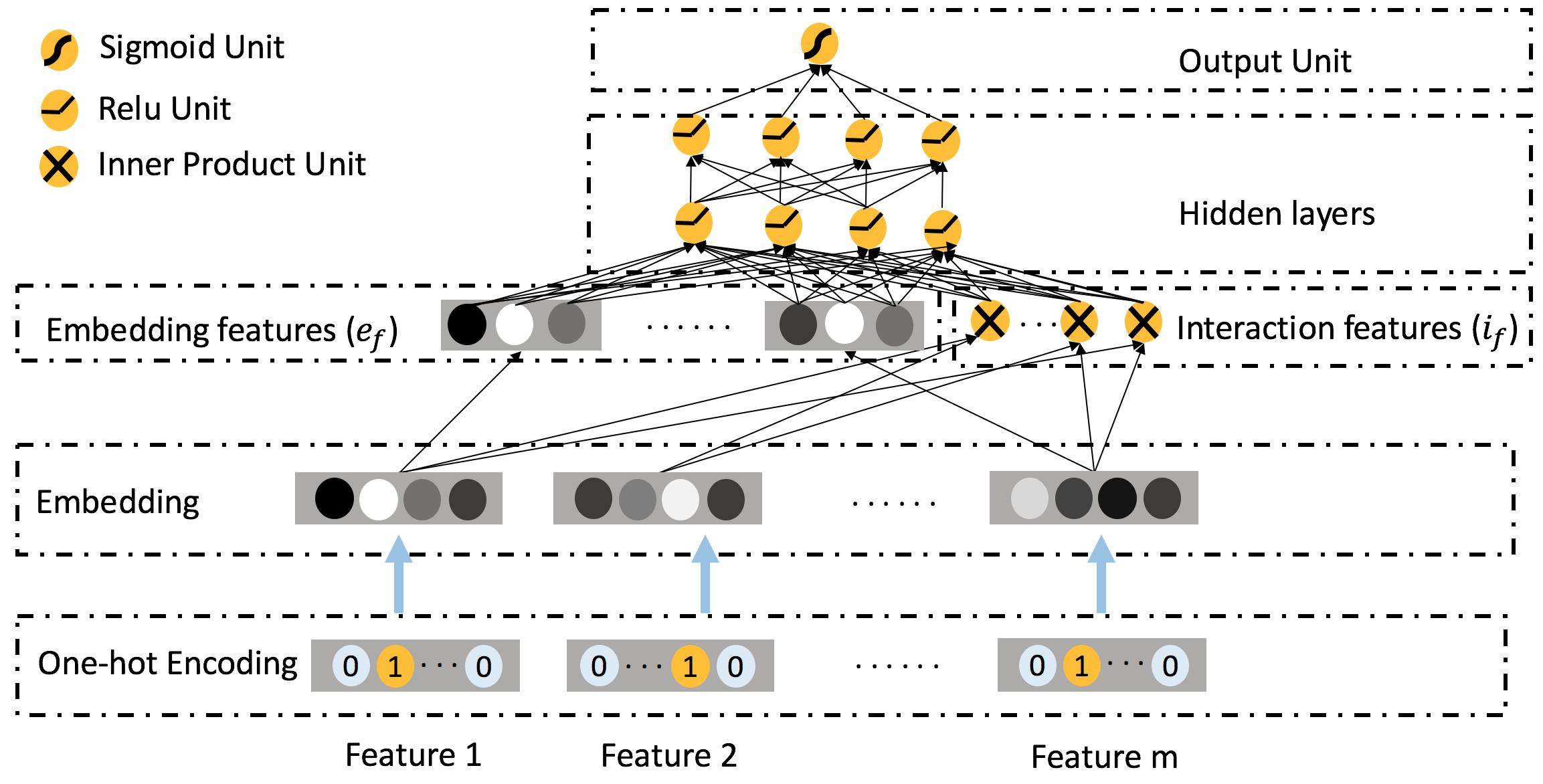}
\caption{The architecture of the PNN model. The structure of PNN is the same as ONN, except that ONN uses operation-aware embedding layer.}
\label{PNN_pic}
\end{figure}

However, FM and FFM lack the ability of mining deep feature dependencies while deep models are good at that.
Via comparing the architectures of FFM and ONN, we observe that the bottom layers of ONN and FFM are the same, but ONN uses several non-linear layers on top of the embedded representations and the product units to mine deep feature dependencies, while FFM directly uses the product units to generate the output.
Thus we say that ONN deepens FFM.
Figure \ref{PNN_pic} illustrates the architecture of PNN proposed in \cite{qu2016product}.
Via comparing the architecture of PNN and FM, it is observable that PNN also deepens FM.


Lastly, via comparing figure \ref{ONN_pic} and figure \ref{PNN_pic}, we observe that the ONN uses the operation-aware embedding, but PNN uses the normal embedding.
Thus ONN makes PNN operation-aware.

Compared with shallow models, deep models can automatically learn sophisticated feature dependencies.
Compared with the normal embedding layer, the operation-aware embedding layer learns better feature representations for different operations.
Thus, ONN is more effective than traditional shallow models and PNN.

\section{Experiments}
In this section, we present our experiments in detail, including datasets, data processing, experimental setups, model comparisons in offline-training setting and online-training setting, and the analyses of different operations. In our experiments, the ONN model outperforms major state-of-the-art models in the CTR estimation task on two real-world datasets in both offline-training and online-training settings.

\subsection{Data}
\paragraph{Criteo}
Criteo dataset\cite{Criteo} includes 45 million users’ click records.
There are 13 continuous features and 26 categorical ones.
We use the last 5 million records for testing and the other records for training.
The continuous features are discretized by the function $discrete(x) = \lfloor2*log(x)\rfloor$, where $\lfloor\cdot\rfloor$ is the floor function.

\paragraph{Tencent Ad} The Tencent Ad dataset\cite{Tencent} is used in the Tencent Social Advertising College Algorithm Competition.
The data contains 14 days app ad conversion data, user information, ad information and the records of installed apps.
We use 39 categorical features which we used in the competition to perform the experiments.
Because the data in the last 2 days is noisy, we use the first 11 days for training, and the 12th day for testing.
After the split, there are about 22 million train data and 2 million test data.

\subsection{Model Comparison}
We compare ONN with 5 models in our experiments, which are implemented with TensorFlow and trained with the Adam optimization algorithm \cite{kingma2014adam}.
\begin{itemize}
\item [FM:] FM learns feature interactions by factorizing it into the inner product of 2 vectors. FM has many successful applications in many CTR/CVR tasks \cite{rendle2010factorization}.
\item [FFM:] FFM is an optimized version of FM. FFM uses field-aware vectors to perform product operation.
\item [DNN:] A simple deep model without product operations. The embedded features are directly fed into an MLP to learn high-order feature interactions and generate output.
\item [PNN:] PNN is proposed in \cite{qu2016product}. PNN introduces the ``inner-product" operation and the ``outer-product" operation into deep model.
In our experiments, the PNN is implemented with ``inner-product" operation, since ONN uses the ``inner-product" operation by default.
\item [DeepFM:] DeepFM trains a deep component and an FM component at the same time \cite{guo2017deepfm}. DeepFM can automatically learn low-order feature interactions and high-order feature interactions at the same time.
\item [ONN:] ONN is the proposed model of this paper.
\end{itemize}
For fairness, we add Batch Normalization(BN) layers to all deep models. The non-linear activation function is set to relu.

\subsection{Evaluation Metrics}
We use four evaluation metrics in our experiments: Area Under ROC(AUC), Cross Entropy(Logloss), Pearson's Correlation Coefficient(Pearson's R) and Root Mean Squared Error (RMSE).
\subsection{Offline-Training Performance Comparison}
We present the performances of models in offline-training environment.
In this setting, the data can be trained several epoches.
Since many hyper-parameters, such as the number of hidden layers, the hidden sizes and embedding dimensionality, have been discussed enough in \cite{qu2016product} and \cite{guo2017deepfm}, we just follow most parameter settings from their works.
The embedding dimensionality is set to 10 for all models.
Although ONN can use flexible embedding dimensionalities, we just use the same embedding dimensionality with other models for fairness.
The number of non-linear hidden layers is set to 3.
For the Criteo dataset, the hidden sizes are set to [400, 400, 400].
For the Tencent Ad dataset, the hidden sizes are set to [200, 200, 200].
Besides, the learning rate of Adam is set by grid search from [0.0001, 0.00025, 0.0005, 0.00075, 0.001] using cross validation.
The training batch size is set to 2500.

\begin{table}[h]
\centering
\caption{Overall Performance on the Criteo dataset in Offline-Training Setting}
\label{poff1}
\begin{tabular}{l|llll}
\hline
\multicolumn{1}{c|}{\multirow{2}{*}{Model}} & \multicolumn{4}{l}{Criteo}  \\ \cline{2-5}
\multicolumn{1}{c|}{}                       & Logloss       & AUC        & Pearson's R & RMSE       \\ \hline
FM                                          & 0.44233       & 0.80464    & 0.48873     & 0.37805    \\
FFM                                         & 0.43846       & 0.80920    & 0.49612     & 0.37627    \\
DNN                                         & 0.43700       & 0.81059    & 0.49924     & 0.37557    \\
PNN                                         & 0.43636       & 0.81134    & 0.50014     & 0.37529    \\
DeepFM                                      & 0.43671       & 0.81150    & 0.49954     & 0.37541    \\
ONN                                         & \textbf{0.43577}& \textbf{0.8123} & \textbf{0.50139} & \textbf{0.37495} \\ \hline
\end{tabular}
\end{table}

\begin{table}[h]
\centering
\caption{Overall Performance on the Tencent Ad dataset in Offline-Training Setting}
\label{poff2}
\begin{tabular}{l|llll}
\hline
\multicolumn{1}{c|}{\multirow{2}{*}{Model}} & \multicolumn{4}{l}{Tencent Ad}  \\ \cline{2-5}
\multicolumn{1}{c|}{}                       & Logloss       & AUC        & Pearson's R & RMSE         \\ \hline
FM                                          & 0.10684       & 0.82376    &  0.26644    & 0.15772  \\
FFM                                         & 0.10639       & 0.82667    &  0.26753    & 0.15767  \\
DNN                                         & 0.10581       & 0.82810    &  0.27261    & 0.15755  \\
PNN                                         & 0.10550       & 0.82725    &  0.27429    & \textbf{0.15734} \\
DeepFM                                      & 0.10595       & 0.82586    &  0.26731    & 0.15768  \\
ONN                                         & \textbf{0.10504} & \textbf{0.82993} & \textbf{0.27481} & 0.15735 \\ \hline
\end{tabular}
\end{table}

Table \ref{poff1} and tabel \ref{poff2} show the overall performance in offline-training setting.
The results show that ONN outperforms all the other model on all metrics.
Besides, the results are consistent with the relationships among FM, FFM, PNN and ONN.
We observe that PNN performs better than FM, and ONN performs better than FFM.
These observations suggest that deep models are more effective than shallow models.
Similarly, we observe that FFM performs better than FM, and ONN performs better than PNN, these observations suggest that the operation-aware embedding is more effective than the normal embedding.
Besides, all of PNN, DeepFM and ONN outperform DNN, suggesting that the ``product" operation is useful in automatically learning feature interactions.

\begin{figure}[ht]
\centering
\includegraphics[width=8.5cm]{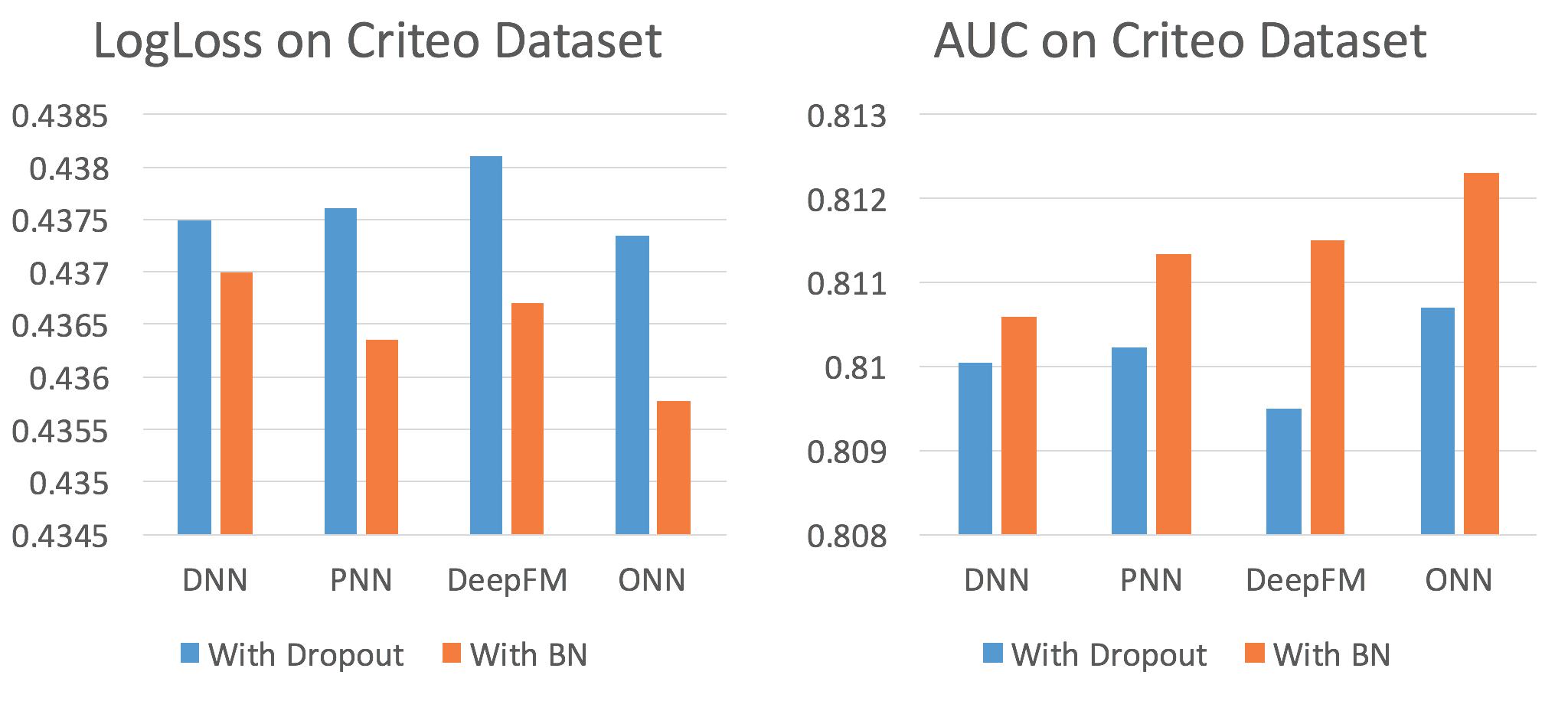}
\caption{Performance Comparison between BN and Dropout.}
\label{BNVSDropout}
\end{figure}

We find that dropout \cite{srivastava2014dropout} does not improve the performance for all deep models after adding BN layer.
In order to make a choice between the two techniques, we compare models with BN layer and models with 0.3 dropout rate on network hidden layers.
Figure \ref{BNVSDropout} illustrates the results.
We observe that BN outperforms dropout for all deep models.
The results suggest that BN is more useful than dropout in deep models for user response prediction tasks.

\subsection{Online-Training Performance Comparison}
In real world online advertising systems, training data are generated in a stream.
Online training is more suitable for this environment than offline training.
In order to validate the effectiveness of deep neural networks in online environment, we compare the performance of different models in online training settings.
In this setting, the data should be trained in a stream.
Mini-batch training is allowed, but the order of data sequence should be kept and each mini-batch can only be learned once in each experiment.
The network parameters are consistent with the offline-training experiments.
But the learning rate is searched again by cross validation.

\begin{table}[ht]
\centering
\caption{Overall Performance on the Criteo dataset in Online-Training Setting}
\label{pon1}
\begin{tabular}{l|llll}
\hline
\multicolumn{1}{c|}{\multirow{2}{*}{Model}} & \multicolumn{4}{l}{Criteo}  \\ \cline{2-5}
\multicolumn{1}{c|}{}                       & Logloss        & AUC       & Pearson's R & RMSE     \\ \hline
FM                                          & 0.44348        & 0.80346   &  0.48695    & 0.37849  \\
FFM                                         & 0.44083        & 0.80643   &  0.49255    & 0.37731  \\
DNN                                         & 0.44132        & 0.80579   &  0.49049    & 0.37763  \\
PNN                                         & 0.44156        & 0.80584   &  0.49005    & 0.37772  \\
DeepFM                                      & 0.44275        & 0.80459   &  0.49033    & 0.37768  \\
ONN                                         & \textbf{0.43751} & \textbf{0.81016} & \textbf{0.49813} & \textbf{0.37578} \\ \hline
\end{tabular}
\end{table}

\begin{table}[ht]
\centering
\caption{Overall Performance on the Tencent Ad dataset in Online-Training Setting}
\label{pon2}
\begin{tabular}{l|llll}
\hline
\multicolumn{1}{c|}{\multirow{2}{*}{Model}} & \multicolumn{4}{l}{Tencent Ad} \\ \cline{2-5}
\multicolumn{1}{c|}{}                       & Logloss       & AUC        & Pearson's R & RMSE          \\ \hline
FM                                          & 0.10721       & 0.81955    &  0.25699    & 0.15817  \\
FFM                                         & 0.10656       & 0.82518    &  0.26572    & 0.15776  \\
DNN                                         & 0.10643       & 0.82584    &  0.26744    & 0.15769  \\
PNN                                         & 0.10698       & 0.82257    &  0.26563    & 0.15776 \\
DeepFM                                      & 0.10653       & 0.82518    &  0.26743    & 0.15775  \\
ONN                                         & \textbf{0.10591}  & \textbf{0.82812}  & \textbf{0.27012} & \textbf{0.15762}  \\ \hline
\end{tabular}
\end{table}

\begin{figure}[h]
\centering
\includegraphics[width=8.5cm]{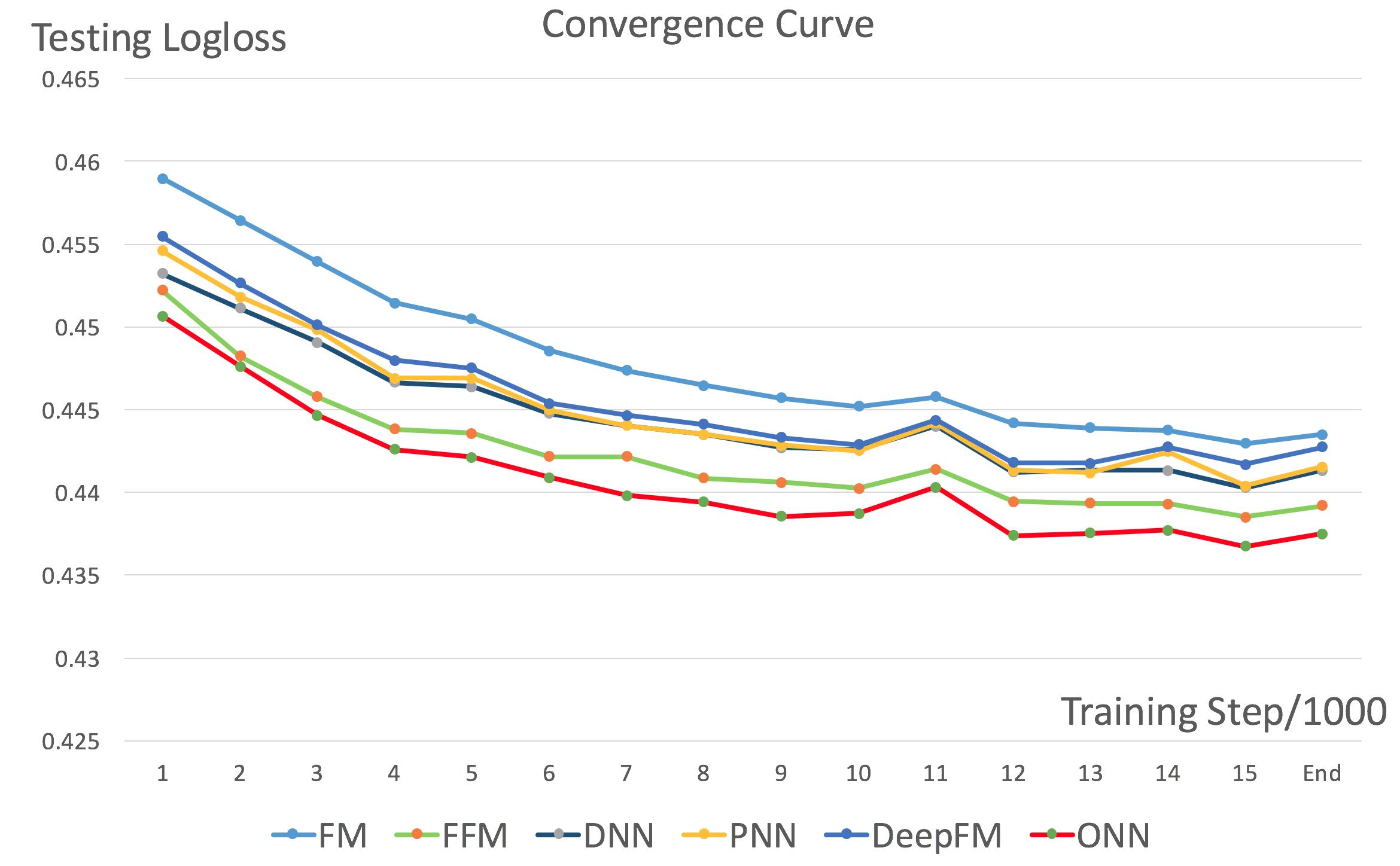}
\caption{Model convergence curves on the Criteo dataset. The horizontal axis is the number of training steps in thousands. The vertical axis is the logloss on the testing dateset.}
\label{CC}
\end{figure}

Table \ref{pon1} and table \ref{pon2} show the overall performance in online-training setting.
The results show that ONN outperforms all the other model on all metrics.
Figure \ref{CC} illustrates the convergence curves of all models on the Criteo dataset. From the figure, we observe that the convergence curve of ONN is always at the bottom among all models, which suggests that ONN converges fastest among all models.

In online-training setting, each sample can only be trained one time.
In FM, PNN and DeepFM, each feature representation needs to compromise among different operations, therefore it is hard for these models to learn a good representation within one epoch.
However, in FFM and ONN which use the operation-aware embedding, each operation has its own representation, thus it is easier for these models to learn feature representations.
Besides, we also observe that DNN outperforms PNN and DeepFM in online-training environment.
We think that is because there is only one ``copy" operation to be done on each feature in DNN, and thus it is easier for DNN to learn feature representations compared with PNN and DeepFM.
After that, ONN outperforms DNN because the ``product" operations make ONN more effective than DNN to learn feature interactions.
In conclusion, multiple operations are useful for deep models, and the operation-aware embedding enables models with multiple operations to learn good feature representations efficiently in online-training environment.

\subsection{ Analysis of Different Operations}
By default, we only use the ``inner-product" operation to learn the feature interactions.
In fact, other operations can also be used in ONN.
In this section, we compare the performance of different ONNs by replacing the default ``inner-product'' operation with ``outer-product'' and ``sub-network'' operations.
The ``sub-network'' operations use a network with a hidden layer to learn the feature interactions between two features.

Besides, we also consider the combination of the ``inner-product'' operation and the ``outer-product'' operation.
We do not consider the combinations of the ``sub-network'' operation with other operations because the space and time complexity of the ``sub-network'' operation is too high, which makes it unsuitable to be used in the real world systems.

\begin{table}[h]
\centering
\caption{Performance comparison for different operations on the Criteo dataset}
\label{pop1}
\begin{tabular}{l|llll}
\hline
\multicolumn{1}{c|}{\multirow{2}{*}{Model}} & \multicolumn{4}{l}{Criteo}  \\ \cline{2-5}
\multicolumn{1}{c|}{}                       & Logloss       & AUC  & Pearson's R & RMSE         \\ \hline
sub-network                                        & 0.43597        & 0.81178      &    0.50081      &  0.37514      \\
outer-product                                         & 0.43673        & 0.81089      &  0.49941        &   0.37547       \\
inner-product                                          & 0.43571        & 0.81016      &   \textbf{0.50188}      &   \textbf{0.37482}      \\
inner+outer-product                                         & \textbf{0.43541}         & \textbf{0.81236}      &  0.50165      &  0.37489      \\ \hline
\end{tabular}
\end{table}

\begin{table}[h]
\centering
\caption{Performance comparison for different operations on the Tencent Ad dataset}
\label{pop2}
\begin{tabular}{l|llll}
\hline
\multicolumn{1}{c|}{\multirow{2}{*}{Model}} & \multicolumn{4}{l}{Tencent Ad}  \\ \cline{2-5}
\multicolumn{1}{c|}{}                       & Logloss       & AUC  & Pearson's R & RMSE         \\ \hline
sub-network                                        & 0.10560        & 0.82612      &  0.27129      &  0.15753      \\
outer-product                                         & 0.10516        & 0.82886      &     0.27328    &    0.15740      \\
inner-product                                          & 0.10504       & 0.82993     &   0.27309       &  0.15744        \\
inner+outer-product                                         & \textbf{0.10502}         & \textbf{0.83028}      &   \textbf{0.27429}      &  \textbf{0.15735}      \\ \hline
\end{tabular}
\end{table}

Table \ref{pop1} and table \ref{pop2} show the experimental results of ONN with different operations.
We can see that the ``inner-product'' operation performs best on the whole among the three single operations.
Besides, we can see that the ``sub-network'' operation also provides competitive performances.
For instance, the ``sub-network'' operation performs best on the AUC metric on the Criteo dataset.
The performances of the``sub-network'' operation indicate that there are more choices other than the ``product'' operations, which can also be used to learn feature interactions in the deep architectures, and we will explore more on that in out future works.
After that, the combination of the ``inner-product'' operation and the ``outer-product'' operation achieves obvious improvement, and performs best among most metrics on the two experimental datasets.

In conclusion, we adopt the ``inner-product'' as the default operation because it performs very competitive and has the lowest space and time complexity.
The ``outer-product'' operation can be used in addition if better performance needs to be achieved at the cost of storage space and computation time.
In fact, the ``outer-product'' operation is much slower than the ``inner-product'' operation.
For the detailed complexity analysis, please refer to \citep{qu2016product}.
\section{Conclusion}
In this paper, we propose a new embedding method named \emph{operation-aware} embedding for learning feature representations in user prediction systems, and construct a new deep neural network named \emph{Operation-aware Neural Networks} (ONN).
Compared with the traditional feature embedding method which learns one representation for all operations, operation-aware embedding can learn various representations for different operations.
Experimental results show that ONN outperforms several state-of-art models on 2 datasets in both offline-training and on-line training environment.
Besides, ONN converges faster than other models and outperforms state-of-art models by a large margin in online-training environment, which suggests that ONN is very suitable for online system.
ONN inherits the main network structure of PNN in this paper, but the operation-aware embedding layer can be applied to any neural architectures actually.

In future, we will explore the use of the operation-aware embedding layer on other applications like NLP.
Besides, we are interested in how to automatically determine the embedding dimensionalities for different operations and different features.
Another question is whether some operations can share one representation.
We will also study on splitting the features and operations into different groups, where the embedded representations are shared in the same group, but not shared among different groups.

\section*{Acknowledgment}
This work is supported in part by the National Science Foundation of China under Grant Nos. (61876076), and Jiangsu NSF grant (BK20141319).

\bibliographystyle{model1-num-names}
\bibliography{onn}

\end{document}